# Hamilton, Hamiltonian Mechanics, and Causation

**Christopher Gregory Weaver**
Assistant Professor of Philosophy, Department of Philosophy[a]
Affiliate Assistant Professor of Physics, Department of Physics[b]
Core Faculty in the Illinois Center for Advanced Studies of the Universe[b]
University of Illinois at Urbana-Champaign

[a]Department of Philosophy
200 Gregory Hall
810 South Wright ST
MC-468
Urbana, IL 61801

[b]Department of Physics
1110 West Green ST
Urbana, IL 61801

**Abstract**: I show how Hamilton's philosophical commitments led him to a causal interpretation of classical mechanics. I argue that Hamilton's metaphysics of causation was injected into his dynamics by way of a causal interpretation of force. I then detail how forces remain indispensable to both Hamilton's formulation of classical mechanics and what we now call Hamiltonian mechanics (*i.e.*, the modern formulation). On this point, my efforts primarily consist of showing that the orthodox interpretation of potential energy is the interpretation found in Hamilton's work. Hamilton called the potential energy function the force-function because he believed that it represents forces at work in the world. Multifarious non-historical arguments for this orthodox interpretation of potential energy are provided, and matters are concluded by showing that in classical Hamiltonian mechanics, facts about the potential energies of systems are grounded in facts about forces. Thus, if one can tolerate the view that forces are causes of motions, then Hamilton provides one with a road map for transporting causation into one of the most mathematically sophisticated formulations of classical mechanics, *viz*., Hamiltonian mechanics.

**Acknowledgements**: An earlier draft of this paper was presented at an online meeting of the California Institute of Technology's philosophy of physics reading group on November 4[th], 2020. I'd like to thank Joshua Eisenthal, Christopher Hitchcock, Mario Hubert, and Charles T. Sebens for their questions and comments on that earlier draft. I thank Michael Townsen Hicks for his comments on an even earlier draft. Special thanks to Tom Banks, Jennifer Coopersmith, and Don Page for helpful correspondence.

"The circular motion is (like every non-rectilinear [motion]) a continuous change of the rectilinear, and as this is itself a continuous change of relation in respect of external space, the circular motion is a change of the change of these external relations in space, and consequently a continuous arising of new motions; since, now according to the law of inertia, a motion, in so far as it arises, must have an external cause, while the body, in every point of this circle, is endeavouring, according to the same law, to proceed in the straight line touching the circle, which motion works against the above external cause, every body in circular motion demonstrates by its motion a moving force…Hence the circular motion of the body…is a *real* motion…a motion cannot take place without the influence of a continuously active external moving force…"[1]

- Immanuel Kant (1724-1804)

"…Very glad you have the Kant from Coleridge for me: try and send it soon. I have read a large part of the *Critique of Pure Reason*, and find it wonderfully clear, and generally quite convincing. Notwithstanding some previous preparation from Berkeley, and from my own thoughts, I seem to have learned much from Kant's own statement of his views of Space and Time. Yet, on the whole, a large part of my pleasure consists in recognizing, through Kant's *works*, opinions, or rather views, which have been long familiar to myself, although far more clearly and systematically expressed and combined by him."[2]

- Sir William Rowan Hamilton (1805-1865)

"All mechanicians agree that reaction is equal and opposite to action, both when one body presses another, and when one body communicates motion to another. All reasoners join in the assertion, not only that every observed change of motion has had a cause, but that every change of motion must have a cause. Here we have certain portions of substantial and undoubted knowledge…We have, in the Mechanical Sciences, certain universal and necessary truths on the subject of causes…Axioms concerning Cause, or concerning Force, which as we shall see, is a modification of Cause, will flow from an Idea of Cause, just as axioms concerning space and number flow from the ideas of space and number or time. And thus the propositions which constitute the science of Mechanics prove that we possess an idea of cause, in the same sense in which the propositions of geometry and arithmetic prove our possession of the ideas of space and of time or number…the relation of cause and effect is a condition of our apprehending successive events, a part of the mind's constant and universal activity, a source of necessary truths; or, to sum all this in one phrase, a Fundamental Idea."[3]

- William Whewell (1794-1866)

"…in Whewell at Cambridge, I thought with delight that I perceived a philosophical spirit more deep and true than I had dared to hope for."[4] "…Whewell has come round almost entirely to my views about the laws of Motion."[5]

- Sir William Rowan Hamilton

---

[1] (Kant 1903, 235-236).

[2] From W.R. Hamilton to Viscount Adare, July 19th, 1834. As quoted by (Graves 1885, 96). The last emphasis is mine.

[3] (Whewell 1858, 182-183).

[4] From W.R. Hamilton to Aubrey De Vere, Observatory, May 7th, 1832. As quoted by (Graves 1882, 554).

[5] From W.R. Hamilton to Lord Adare dated April, 1834. As quoted in (Graves 1885, 83).

## Hamilton, Hamiltonian Mechanics, and Causation

## 1    Introduction

Sir William Rowan Hamilton (1805-1865) affirmed that every dynamical evolution is a causal evolution. His causal dynamics may seem out of place to the contemporary philosopher of physics because most everyone in that subdiscipline now agrees with David Papineau's remark that "there is strong reason to doubt that causation is constituted by basic dynamical processes".[6] These same philosophers also typically agree with Michael Redhead's (1929-2020) comment that "physicists long ago gave up the notion of cause as being of any particular interest! In physics, the explanatory laws are laws of functional dependence…"[7] Alongside this majority opinion is a somewhat prevalent and friendly (to the majority opinion) story about causation and the historical development of physics and its supporting mathematics, a story alluded to by Redhead. It says that as physics grew more and more mathematically sophisticated, less and less room remained for causal relations in the ontologies of our best physical theories. As Marius Stan concluded very recently:

> …the basic laws no longer count as causal principles after 1790; nor do their local corollaries, the equations of motion. In effect, the laws no longer state facts about the basic causal powers of bodies, material substances, or their ontological analogues…This astonishing consequence—it really heralds the disappearance of causality from fundamental science—remained unnoticed until Russell came across it. Disappointingly for a modern reader, it eluded the 18th century's great minds. Oblivious to the tectonic shift beneath their feet, philosophers and reflective men of science alike continued to refer to physics, whose rock bottom was then mechanics, as the 'science of causes.' But a hard-headed interlocuter might wonder where causes were hidden, seeing as they no longer reside in the laws.[8]

I will argue that this conclusion (and so the prevalent story) is multiply flawed.[9] As I've already said, and as I will argue in **sect. 2**, Hamilton's (to say nothing of Whewell's) *19th century* mechanics was understood by him to be a causal approach to that mechanics, and, what appears to be all the rage these days notwithstanding, Hamilton's two most important and revolutionary papers on classical mechanics (Hamilton 1834; Second Essay 1835) provide persuasive ammunition for a causal interpretation of the laws of *modern* classical physics. The student of the history of classical physics knows just how important these two papers are. Indeed, it is hard to

---

overestimate their value, for in them, Hamilton impacted our modern understanding of the underlying mathematics of classical mechanics to the extent that when introducing classical mechanics to students, modern physics textbooks typically present Hamiltonian mechanics, that modern and sophisticated formulation of classical mechanics inaugurated by (Hamilton 1834, Second Essay 1835).[10]

## 2    Hamilton's Metaphysical Dynamics
### 2.1    Influences and Methodology: A Sample

In his monumental work, *Critique of Pure Reason*, Kant presented the "Second Analogy of Experience". According to that analogy, any and all transmutation or change must be accompanied by obtaining causal relations.[11] As Kant's "principle of temporal sequence according to the law of causality" (a synthetic *a priori* judgment) stated, "[a]ll alterations occur in accordance with the law of the connection of cause and effect."[12] For Kant, every substance undergoing change is connected to something else. In 1755, some 26 years before the first *Critique*, Kant published the *New Elucidation*. The last section of that work explicates Kant's Principle of Succession, which says that:

> No change can happen to substances except in so far as they are connected with other substances; their reciprocal dependency on each other determines their reciprocal change of state.[13]

The connection with which Kant is concerned is the connection between cause and effect. That relation is not reducible to Humean regularity or constant conjunction (see Friedman, Causal Laws 1992, 162; Kant 1933, 4-6). For Kant, the connection in succession bespeaks a modal tie because causation is a modal tie indicative of the natural necessitation of an effect by its cause. As Kant remarked in the *Prolegomena to Any Future Metaphysics* (1783):

> For this concept [of causation] positively requires that something A be such that something else B follow from it *necessarily* and *in accordance with an absolutely universal rule*…the effect is not merely joined to the cause, but rather is posited *through* it and results *from* it.[14]

---

[10] As Jos Uffink has written, "I would remark in defense of textbook writers anonymous, that they, usually discuss Hamiltonian mechanics rather than Newtonian." (Uffink 2002, 557)

For a thorough study of the development of Hamilton's mechanics, one should begin with (Hamilton 1833).

[11] (Kant 1998, 304-316; B232-B258) where the latter reference cites in accordance with the section divisions of the 1787 edition of the first *Critique*.

[12] (Kant 1998, 304; B232). I have removed the emphasis of the first quotation in this sentence.

[13] (Kant 1992, 37). For Kant, this is a principle of "metaphysical cognition" (ibid.). Its tincture reminds one of Isaac Newton's (1643-1727) third law of motion (*i.e.*, the action-reaction principle). In the *Prolegomena to Any Future Metaphysics*, Kant annunciated three *a priori* principles about the cognition of appearances in experience. The last of these invokes an action-reaction principle (reminiscent of Newton's third law of motion and the reciprocity mentioned in the quotation in the main text) understood as a type of epistemic guide or tool (Kant 1933, 66).

[14] I quote here the translation of Kant's *Prolegomena to Any Future Metaphysics* (A91-92/B123-124) provided by Michael Friedman in (Friedman, Causal Laws 1992, 161-162) emphasis in the original. For a standard English translation of Kant's *Prolegomena*, see (Kant 1933) or (Kant 2004).

Compare: "the concept of cause implies a rule, according to which one state follows another necessarily" (Kant 1933, 76).

Universal and naturally necessary rules or laws back causal relations and are for Kant, therefore, causal laws.[15] Elsewhere in the same work Kant would add:

> …natural necessity must be the condition on which effective causes are determined…If without contradiction we can think of the beings of understanding as exercising such an influence on appearances, then natural necessity will attach to all connexions of cause and effect in the sensuous world...[16]

The excerpt reveals how Kant's opinions about causation were not left isolated in his theoretical philosophy. They entered his natural philosophy as well. The laws are causal precisely because they reference quantities that ought to be interpreted as producers of distinctive effects. For example, a body accelerated from (relative) rest to some subsequent uniform motion enjoys an alteration of state. That alteration requires a cause, and that cause, for Kant is a force that naturally necessitates its effect. A second example comes from Kant's own work. When a body enjoys a curvilinear or circular motion that motion is due to a cause that is a dynamically acting force bringing about a continual alteration of rectilinear motion. As Kant put it in his *Metaphysical Foundations of Natural Science* (1786):

> The circular motion is (like every non-rectilinear [motion]) a continuous change of the rectilinear, and as this is itself a continuous change of relation in respect of external space, the circular motion is a change of the change of these external relations in space, and consequently a continuous arising of new motions; since, now according to the law of inertia, a motion, in so far as it arises, must have an external cause, while the body, in every point of this circle, is endeavouring, according to the same law, to proceed in the straight line touching the circle, which motion works against the above external cause, every body in circular motion demonstrates by its motion a moving force…Hence the circular motion of the body…is a *real* motion…a motion cannot take place without the influence of a continuously active external moving force…[17]

Hamilton's natural philosophy was greatly influenced by Immanuel Kant's theoretical and natural philosophy.[18] Kant's Second Analogy of Experience led Hamilton to the view that physical

---

[15] There is a question about whether Kant's laws of mechanics are identical to Newton's laws of motion. For the view that they are identical, see (Friedman 2013). For the view that they are distinct, but perhaps similar, see (Watkins 2005). I should add that Watkins's views have undergone an evolution. Compare (ibid.) to (Watkins 2019).

[16] (Kant 1933, 111-112).

[17] (Kant 1903, 235-236). See also (Friedman 2013, 335-412).

[18] By 1834, Hamilton would express admiration for Kant in correspondence with Viscount Adare:

> "…Very glad you have the Kant from Coleridge for me: try and send it soon. I have read a large part of the *Critique of Pure Reason*, and find it wonderfully clear, and generally quite convincing. Notwithstanding some previous preparation from Berkeley, and from my own thoughts, I seem to have learned much from Kant's own statement of his views of Space and Time. Yet, on the whole, a large part of my pleasure consists in recognizing, through Kant's *works*, opinions, or rather views, which have been long familiar to myself, although far more clearly and systematically expressed and combined by him." From W.R. Hamilton to Viscount Adare, Observatory, July 19th, 1834. As quoted by (Graves 1885, 96). The last emphasis is mine.

evolutions must involve obtaining causal relations,[19] and it would even help Hamilton justify a new interpretation of algebra.[20] However, it was the former influence relation—the one resulting in Hamilton's *Kantian* conviction that dynamics must be interpreted causally—that encouraged him to make room for causally efficacious forces in the ontology of his new formulation of mechanics (*q.v.*, **sect. 4.1**).

To discern even more clearly how Hamilton ensured that his mechanics remained committed to a causal ontology, look beyond the influence of Kant to see how Hamilton viewed his work in relation to that of Joseph-Louis Lagrange (1736-1813), a brilliant mathematician and physicist responsible for the initial development of a formulation of classical mechanics known today as Lagrangian mechanics. Hamilton said that mechanics is the "the science of force, or of power acting by law in space and time" and that it has

> …undergone already another revolution and has become already more dynamic, by having almost dismissed the conceptions of solidity and cohesion, and those other material ties, or geometrically imaginable conditions, which Lagrange so happily reasoned on, and by tending more and more to resolve all connexions and actions of bodies into attractions and repulsions of points…[21]

For Hamilton, mechanics is the study of forces acting by means of repulsions or attractions on point masses situated in space and time.[22] These forces are the causes of motions or changes thereof. It is not a coincidence that Lagrange's characterization of dynamics in his *Analytical Mechanics* (1788) is remarkably close to Hamilton's:

> Dynamics is the science of accelerating or retarding forces and the diverse motions which they produce…the discovery of the infinitesimal calculus enabled geometers to reduce the laws of motion for solid bodies to analytical equations and the research on forces and the motion which they produce has become the principal object of their work.[23]

When presenting his own methodology for the study of optics, Hamilton (1931, 314) approvingly quotes Newton's *Opticks*. The passages quoted explicate Newton's theory of analysis or composition. Newtonian analysis begins by empirical observation and experimentation. It then

---

There are reasons to believe that Hamilton was one of the first British theoreticians to seriously study Kant. See (Hankins 1980, 179) who notes that William Whewell (1794-1866) also helped spread Kantian ideas to British thinkers. I can find hardly any work on the relationship between Kant and Hamilton.

[19] (Hankins 1980, 178-179).

[20] As Graves's summary indicates,

> "…we regard him [Hamilton] as having made a decidedly Kantian movement, when he conceived and published that view of algebraic science, including the various calculi, to which we have just now referred." (Graves 1842, 107)

See also (Graves 1885, 141-143). For more on Hamilton and algebra, see (Merzbach and Boyer 2011, 510-512).

[21] (Hamilton 1834, 247).

[22] Elsewhere, he says that dynamics, or what he called "Dynamical Science", consists of "reasonings and results from the notion of cause and effect" (Hamilton, Conjugate Functions 1835, 7).

[23] (Lagrange 1997, 169).

recommends cogent inductive inferences to general conclusions on the basis of experimental results and observations. Hamilton records Newton's remarks as follows:

> By this way of analysis, we may proceed from compounds to ingredients, and from motions to the forces producing them; and, in general, from effects to their causes, and from particular causes to more general ones, till the argument end in the most general. This is the method of analysis: and the synthesis consists in assuming the causes discovered, and established as principles, and by them explaining the phenomena proceeding from them, and proving the explanations…[24]

For good measure, Hamilton adds (after his quotation of Newton) that "[i]n the science of optics, which has engaged the attention of almost every mathematician for the last two thousand years, many great discoveries have been attained by" means of utilizing the above Newtonian methodology.[25] Hamilton's reliance upon Newton for an explication of his own scientific method supports the judgment that Hamilton believed that forces cause motions, and that scientists engaged in sound physical inquiry investigate the natural world with the aim of uncovering causal structure revealed in the actions of forces.[26] Forces are the means whereby causation enters classical mechanics.

Hamilton wasn't just influenced by Kant, Lagrange, and Newton. He also had numerous and important interactions with the great polymath William Whewell (1794-1866).[27] Whewell was an immensely important intellectual in the 19th century. His influence on the development of physics can not only be seen in his central role in the invention of terms like 'physicist', 'anode', 'cathode', and 'electrolysis', but it can also be seen in the impact he left on the thought of Michael Faraday (1791-1867), James Clerk Maxwell (1831-1879), and, of course, Hamilton.[28]

Whewell wrote extensively on classical mechanics. He even articulated a unique formulation and interpretation of the three classical laws of motion and sought to justify those laws with their accompanying interpretations by appeal to an elaborate metaphysics of causation.[29] Like Newton[30] and Lagrange[31], Whewell believed that forces cause motions. He said that "the term *Force*" denotes "that property which is the cause of motion produced, changed, or prevented,"[32] and that:

---

[24] I quote Hamilton's exact quotation of Newton at (Hamilton 1931, 314). But see (Newton 1952, 404-405).

[25] (Hamilton 1931, 315).

[26] Hamilton's language elsewhere is quite strong. He believed in axioms of causation which entail that bodies in motion, particularly those moving along curves, enjoy that motion because of causes that produced that motion (Hankins 1980, 178). By using axioms of causation, Hamilton may have been showing the influence of Whewell on his work.

[27] On Whewell, see (Butts 1968); (Fisch and Schaffer 1991); (Whewell 1836, 1858, 1967) and for valuable correspondence, see (Todhunter, vol. 1 and vol. 2 1876).

[28] For a discussion of Whewell's influence on Maxwell, see (Smith 1998, 305). For a discussion of Whewell's influence on Faraday, see (Darrigol 2000, 83-86, 97). For a discussion of Whewell's influence on Hamilton, see (Hankins 1980, 172-180) on which I lean in part.

[29] See (Whewell 1836, 138-161); (Whewell 1967, 573-594).

[30] Newton said, "…the forces…are the causes and effects of true motions." (Newton 1999, 414, cf. 407, 575, 794). See also (Dobbs 1992, 207-209); (M. Jammer 1957); (McGuire 1968); (McGuire 1977); (B. Pourciau 2006, 188-189); (R. Westfall 1971); and *q.v.*, n. 24.

[31] Lagrange said, "[i]n general, **force** or **power** is the cause, whatever it may be, which induces or tends to impart motion to the body to which it is applied." (Lagrange 1997, 11) emphasis in the original.

[32] (Whewell 1858, 205) emphasis in the original.

> By Cause we mean some quality, power, or efficacy, by which a state of things produces a succeeding state. Thus the motion of bodies from rest is produced by a cause which we call Force: and in the particular case in which bodies fall to the earth, this force is termed Gravity. In these cases, the Conceptions of Force and Gravity receive their meaning from the Idea of Cause which they involve: for Force is conceived as the Cause of motion.[33]

Hamilton and Whewell maintained an important and mutually beneficial professional relationship. And while I cannot now explore all of the details of their interactions (see Graves 1882, 1885, 1889; Hankins 1980, 174-180), I note here that the two seemed to agree on how best to interpret the laws of motion. *After* authoring (Hamilton 1834)[34], Hamilton sent a letter (dated March 31[st], 1834) to Whewell in which he stated:

> The Paper of your own On the Nature of the Truth of the Laws of Motion has been as yet so hastily read by me, that I can only say it seems to be an approach, much closer than of old, between your views and mine. Whether this approach is a change on my part or on yours, and if both, in what proportion, and how much or how little it wants of a perfect agreement, I dare not suddenly decide.[35]

The following month, Hamilton wrote to Lord Adare as follows: "Whewell has come round almost entirely to my views about the laws of Motion."[36] Hamilton believed that the interpretation of the laws of motion in Whewell's "On the Nature of the Truth of the Laws of Motion"[37] (published in 1834) matched his own interpretation of those laws. In that essay, Whewell said:

> The science of Mechanics is concerned about motions as determined by their causes, namely, forces; the nature and extent of the truth of the first principles of their science must therefore depend upon the way in which we can and do reason concerning *causes*.[38]

Whewell went on to state what he calls "Axiom 1": "*Every change is produced by a cause*."[39] Axiom 2 says that: "*Causes are measured by their effects*."[40] Axiom 3 was about action and reaction. These axioms are used to motivate and interpret the laws of motion. The ensuing discussion also very clearly and explicitly and continually interprets forces causally, characterizing them, again and again, as causes of motion.[41] Because Hamilton agreed with the views of Whewell here articulated, we can safely conclude that Hamilton adopted a causal metaphysics of physics amidst the two years he published his two most important papers on classical mechanics.[42]

---

[33] (Whewell 1858, 173). See also the comments at (Smith and Wise 1989, 362).

[34] The paper was received by *Philosophical Transactions of the Royal Society* on April 1[st], 1834 and read April 10[th], 1834. However, at the end of Hamilton's introductory remarks, Hamilton includes a date of March, 1834, and because his letter to Whewell references (Hamilton 1834 although it was not yet published) we can infer that Hamilton's letter (dated March 31[st], 1834) was authored after Hamilton had completed but not yet published (Hamilton 1834).

[35] From W.R. Hamilton to Dr. Whewell, Observatory, Dublin, March 31, 1834 in (Graves 1885, 82).

[36] As quoted in (Graves 1885, 83).

[37] (Whewell 1967, 573-594).

[38] (Whewell 1967, 574) emphasis in the original.

[39] (Whewell 1967, 574) emphasis in the original.

[40] (Whewell 1967, 575) emphasis in the original.

[41] The idea is all over the essay, but see (ibid., 581) for just one (more) example among many.

## 2.2    Hamilton's Idealism: A Sample

Hamilton was an idealist about the natural world. As I have already pointed out, Hamilton's views about the role of causation in dynamics were heavily influenced by Kant's first *Critique*. It is therefore unsurprising to see the influence of the first *Critique* on Hamilton's idealism as well. There are also important similarities between Hamilton's idealism and philosophical worldview, and the idealism and worldview of Gottfried Wilhelm Leibniz's (1646-1716) *Monadology* (Leibniz 1898 originally published in 1714). Leibniz maintained that there are scientific efficient causal explanations of motion and change in the kingdom of power (science/physics) and divine final causal explanations of everything in the kingdom of wisdom (philosophy/theology).[43] There is something approaching this bifurcation in the work of Hamilton, for in that work, one finds efficient causal scientific explanations divorced from his idealism on the one side, and deeper more metaphysical explanations dependent upon his idealism but ultimately bottoming out in God on the other.[44] There were for Hamilton (quoting Hankins) "two separate sciences joined only by the benevolent act of God."[45]

In the kingdom of wisdom, Leibniz had fundamental simple monads metaphysically explain a great deal. Hamilton had his fundamental and simple energies or powers play a similar role in the Hamiltonian worldview.[46] As with Hamilton's indebtedness to Kant, a thorough study explicating all of the important similarities and differences should be pursued, but I will not take up that task here. I wish only to disclose that Hamilton's fundamental and simple energies possess *causal powers*. Indeed, sometimes Hamilton speaks as if the fundamental energies are identical to causal powers acting in space and time (Hankins 1977, 179). As Hamilton said, "[p]ower, acting by law in Space and Time, is the ideal base of an ideal world, into which it is the problem of physical science to refine the phenomenal world…"[47]

---

## 3        A (Very) Brief Sketch of Modern Hamiltonian Mechanics

Hamilton's methodology, intellectual influences, and philosophical worldview all led him to a causal interpretation of classical mechanics. Forces in that mechanics causally produce changes of motion. The modern philosopher of physics will object. Nowhere in the fundamental equations of *contemporary* Hamiltonian mechanics (at least) do we see an indispensable role for forces. Hamilton's views are his own. How could, and why should his ideas influence our modern understanding? Contemporary Hamiltonian mechanics is an *energy*-based theory which seeks to describe and explain mechanical systems by means of the Hamiltonian $\mathcal{H}$. $\mathcal{H}$ is equal to the sum of the kinetic $T$ and potential energy $U$ of the modeled system when, potential energy is velocity independent and what are called generalized coordinates (introduced below) are natural in the sense that their relationship to the relevant rectangular Cartesian coordinates is time-independent (Marion and Thornton 1988, 218-219). And so, when assuming such naturalness of coordinates and when assuming that the target system (call it SYS) is an energetically isolated holonomic *n*-particle system, it follows that:

$$\mathcal{H} = T + U$$

and that the number of generalized coordinates and the number of degrees of freedom of SYS equal one another.

According to the Hamiltonian formulation, the configuration of SYS (in three-dimensions) is given by *n*-generalized coordinates: $(q_1, \dots, q_n)$. Following (Goldstine, Poole and Safko 2002), (Marion and Thornton 1988, 222-229), (Taylor 2005, 529-531), and (Thorne and Blandford 2017, 158-160) one can represent *n*-generalized coordinates with $\mathbf{q}$, and generalized momenta[48] with $\mathbf{p}$ such that,

$$\mathbf{q} = (q_1, \dots, q_n), \ \ \mathbf{p} = (p_1, \dots, p_n)$$

We can represent generalized velocities, or the time derivatives of the generalized coordinates with $\dot{\mathbf{q}}$, such that,

$$\dot{\mathbf{q}} = (\dot{q}_1, \dots, \dot{q}_n)$$

Quantities $\mathbf{p}$ and $\mathbf{q}$ above are *n*-dimensional vectors in abstract spaces (Taylor 2005, 529).[49] Assuming that $(i = 1, \dots, n)$ and, when applicable, that $(j = 1, \dots, m)$ here and throughout, one can write Hamilton's equations (of motion) as:

---

[48] All discussion of modern physics will use the SI unit system.

Generalized momenta can also be stated in terms of the Lagrangian $\mathcal{L}$ (*q.v.*, n. 50 below) and generalized velocity (Penrose 2007, 476, I'm citing in this case because some suggest otherwise):

$$\mathbf{p} = \frac{\partial \mathcal{L}}{\partial \dot{q}}$$

where $\mathcal{L}$ is the Lagrangian and *not* the Lagrangian density.

[49] In technical discussions of Hamiltonian mechanics in the work of physicists, mathematicians, and some philosophers, one will see: (a) higher-dimensional phase spaces the points of which represent possible states of the system modeled (because they encode information about the positions and momenta of constituents of the system), (b) phase space orbits or flows tracing out (c) curves in phase space understood as representations of possible

(**Hamilton's Equations of Motion**):

$$\dot{p}_i = -\frac{\partial \mathcal{H}}{\partial q_i}, \quad \dot{q}_i = \frac{\partial \mathcal{H}}{\partial p_i}$$

Add to the above a specification of the (full) time derivative of the Hamiltonian as follows:

$$\frac{d\mathcal{H}}{dt} = \sum_{i=1}^{n}(\frac{\partial \mathcal{H}}{\partial q_i}\dot{q}_i + \frac{\partial \mathcal{H}}{\partial p_i}\dot{p}_i) + \frac{\partial \mathcal{H}}{\partial t}$$

And keep in mind that the full or total derivative of the Hamiltonian with respect to time features $2n$+1 terms (as can be read off of the equation just stated) and that:

$$\frac{d\mathcal{H}}{dt} = \frac{\partial \mathcal{H}}{\partial t}$$

which can be inferred from the equations of motion and our specification of the full time derivative of the Hamiltonian.[50] And while the above expressions equal one another, they are *conceptually* distinct. The left side describes how the Hamiltonian changes with time as the generalized momenta and coordinates change with time. The right side describes the rate of change of the Hamiltonian as time changes, but while holding its "other arguments fixed" (Taylor 2005, 530). The equality will hold for systems like SYS because the Hamiltonian describing them is a constant of motion.

Solutions to Hamilton's equations give one the evolution of the system or subsystems modeled. The contemporary philosopher of physics will point out that forces do not seem to enter any of the above relations or equations. Likewise, forces do not appear to enter solutions to

---

evolutions of the modeled system given by solutions to Hamilton's equations, (d) Liouville's theorem, (e) measures, (f) Poisson brackets, *etc*. For all of that, see (Dürr and Teufel 2009, 12-26); (Mann 2018, 167-201); (Torres del Castillo 2018, 103-228) and pair it with (Healey 2007, 248-251). I skip that stuff here in the interest of brevity. My central argument will remain unaffected by details about cotangent bundles or phase spaces that are symplectic manifolds, measure preserving flows, and symplectic geometry. What gives you the curves that represent evolutions in the phase space are solutions to Hamilton's equations. So, the question is, how should one interpret those equations and their solutions?

Later I will make much of Galilean invariance in classical mechanics. Some might therefore object to precluding a discussion of the geometry of Hamiltonian mechanics because both canonical transformations and canonical invariants (or canonical form-invariants) are important to Hamiltonian and Hamilton-Jacobi mechanics. In order to appreciate canonical transformations (especially those that have to do with time), one must study symplectomorphisms and that study will require that one give attention to cotangent bundles and symplectic geometry. Canonical transformations have to do with tracking systems in a higher-dimensional phase space. Hamilton's equations of motion are canonical form-invariant, and (again) their solutions provide one with the motions of systems modeled by the Hamiltonian apparatus (points orbiting in phase space). Once again, the question is, how should we interpret those equations and their solutions?

[50] Hence the logical ordering of the equations in the main text. We can also affirm:

$$-\frac{\partial \mathcal{L}}{\partial t} = \frac{\partial \mathcal{H}}{\partial t}$$

Hamilton's equations of motion. How then can the contemporary natural philosopher insist on a causal interpretation of Hamiltonian mechanics? Indeed, if Hamilton's understanding was anything like our own, how could *Hamilton* have done so?

As we shall see, there are powerful reasons for insisting that forces belong to the ontology of modern Hamiltonian mechanics. Those reasons shall issue forth from reflecting upon precisely why those same forces never left the ontology of *Hamilton's* mechanics despite the privileged place of the Hamiltonian in his formulation. The crucial piece to my argumentation will actually turn out to be the Hamiltonian. A proper interpretation of that quantity along with a correct statement of its metaphysical grounds (according to physics!) provides much of what is needed for the articulation and defense of a robust case for injecting forces into both past and present versions of Hamiltonian mechanics.

## 4    Hamiltonian Causal Mechanics
### 4.1    Hamilton's Dynamics: A Brief Exposition

In (Hamilton 1834), the **law of varying action** for SYS (*i.e.*, a system of $n$-point masses with an $i^{\text{th}}$ member[51]) is stated as follows:

**(Law of Varying Action (LVA)**[52]):

$$\delta V = \sum_{i=1}^{n} m_i \left( \dot{x}_i \delta x_i + \dot{y}_i \delta y_i + \dot{z}_i \delta z_i \right) - \sum_{i=1}^{n} m_i \left( \dot{a}_i \delta a_i + \dot{b}_i \delta b_i + \dot{c}_i \delta c_i \right) + t \delta \mathcal{H}$$

where $m$ is mass, $t$ is time, $x_i$, $y_i$, and $z_i$ are Cartesian coordinate variables, $a_i$, $b_i$, and $c_i$ give the initial positions of the $n$-point masses, and $\dot{a}_i$, $\dot{b}_i$, and $\dot{c}_i$ give the initial velocities of the same point-masses with regard to select coordinate directions, and so the dots are derivatives with respect to time. The '$\delta$' symbol was originally invented by Lagrange to track some of the maneuvers of Leonhard Euler (1707-1783). Later on, in both Lagrange and Hamilton, it was used to represent a variation or specific type of differential change of a function or functional, although (in Lagrange's work at least) it can act on operators as well.[53] $\mathcal{H}$ is now such that:

$$\mathcal{H} = T + -U$$

because I *now* assume with Hamilton that $U$ is "the negative of the potential energy".[54] (More on this function below.) Hamilton picks out the sum with the letter '$\mathcal{H}$' in honor of Christiaan Huygens (1629-1695).[55] For SYS, $T$'s value is given by what Hamilton called "the celebrated law of living forces" (*q.v.*, n. 56):

---

(**Law of Living Force (LLF)**[56]):

$$T = \mathcal{H} + U$$

In addition:

$$T = \frac{1}{2}\sum_{i=1}^{n} m_i \left( \dot{x_i}^2 + \dot{y_i}^2 + \dot{z_i}^2 \right)$$

All of this is just to make explicit what we've already noted, SYS is conservative.

The variation on $T$ is expressed in this context as:

$$\delta T = \delta U + \delta \mathcal{H} = \sum_{i=1}^{n} m_i \left( \dot{x}_i \delta \dot{x}_i + \dot{y}_i \delta \dot{y}_i + \dot{z}_i \delta \dot{z}_i \right)$$

$V$ in **LVA** is what Hamilton called the "action of the system", the "characteristic function", and the "accumulated living force".[57] $V$ is a function of $\mathcal{H}$ along with $x, y, z, a, b$, and $c$. It must satisfy, according to Hamilton, the following two fundamental partial differential equations:

(**Constraining Fundamental PDEs**):

$$\begin{cases} \dfrac{1}{2}\displaystyle\sum_{i=1}^{n} \dfrac{1}{m_i}\left\{ \left(\dfrac{\partial V}{\partial a_i}\right)^2 + \left(\dfrac{\partial V}{\partial b_i}\right)^2 + \left(\dfrac{\partial V}{\partial c_i}\right)^2 \right\} = T_{t_0} \\ \dfrac{1}{2}\displaystyle\sum_{i=1}^{n} \dfrac{1}{m_i}\left\{ \left(\dfrac{\partial V}{\partial x_i}\right)^2 + \left(\dfrac{\partial V}{\partial y_i}\right)^2 + \left(\dfrac{\partial V}{\partial z_i}\right)^2 \right\} = T \end{cases}$$

where the top constraining fundamental PDE gives the living force (kinetic energy) of the system at the start of the evolution, and where the bottom equation gives the living force of the system at the end. Hamilton (1834, 253) says that one must use these equations to retrieve the form of the characteristic function $V$.

To see what **LVA** means one need only understand the content and significance of the characteristic function and a little calculus of variations. Function $V$ (quoting Hankins) "completely determines the mechanical system [SYS] and gives us its state at any future time once the initial conditions are specified".[58] It is therefore *the law* which governs the evolutions of mechanical systems of the same type as SYS. The **LVA** was a central *tool* that Hamilton used to build his dynamics. It is also central to my case for the historical judgment that the dynamical models of (Hamilton 1834) and (Hamilton, Second Essay 1835) were correctly causally interpreted by Hamilton. Thus, reflection upon the **LVA** will reveal precisely how causation entered

---

[56] (Hamilton 1834, 250).
[57] (Hamilton 1834, 251-252). Hamilton had already written about the characteristic function in (Hamilton 1828) at the age of 21. According to Sir Edmund Whittaker (1873-1956), Hamilton discovered the function at the age of 16 (Whittaker 1954, 82).
[58] (Hankins 1980, 186).

Hamilton's dynamics and by consequence it will also reveal how causation enters modern Hamiltonian mechanics.

The adroit reader has probably already noticed that I have failed to explicate the meaning and significance of one choice function in both the **LLF** and (indirectly via the Hamiltonian) the **LVA**, *viz.*, the potential energy function $U$. Concerning that function, Hamilton wrote, "[t]he function which has been here called $U$, may be named the *force-function* of a system…it is of great utility in theoretical mechanics, into which it was introduced by Lagrange…"[59] Hamilton's indebtedness to Lagrange becomes even more evident after reflecting upon the fact that Hamilton specified the force-function's variation by means of the dynamical principal of virtual work:

(**Dynamical Principle of Virtual Work (DPVW)**):

$$\delta U = \sum_{i=1}^{n} m_i \left( \ddot{x}_i \delta x_i + \ddot{y}_i \delta y_i + \ddot{z}_i \delta z_i \right)$$

which is basically Lagrange's **principle of virtual work (PVW)** (sometimes called d'Alembert's principle).[60] The variation of the potential energy function represents changes to work performed over time by the point masses of SYS, because (at least in part) potential energy represents work performed by the constituents of SYS.[61] Thus, Hamilton affirms

(**Force-Function (FF)**):

$$U = \sum_{i,j=1}^{n,m} m_i m_j f\left(r_{ij}\right)$$

---

[59] (Hamilton 1834, 249) emphasis in the original. It was very common during Hamilton's time to call potential energy the force-function (Nakane and Fraser 2002, 163). As a prime example, Nakane and Fraser mention Carl Gustav Jacob Jacobi (1804-1851) who would, in some ways, improve Hamilton's work. Interestingly, Leonhard Euler (1707-1783) identified what Lagrange called the potential as force *effort* (Euler 1753, 173-175); (Boissonnade and Vagliente 1997, xxxvii); (Weaver draft).

[60] As in (Weaver draft), here my presentation is indebted to the presentations in (Hankins 1980, 184-186) and (Nakane and Fraser 2002, 163). Lagrange's **PVW** in modern vector notation would read as follows:

(**Principle of Virtual Work (PVW)**):

$$\sum \mathbf{F} \cdot \delta \mathbf{r} = \sum m\ddot{\mathbf{r}} \cdot \delta \mathbf{r}$$

Or:

(**Modified Modern Principle of Virtual Work (MMPVW)**):

$$\sum m\ddot{\mathbf{r}} \cdot \delta \mathbf{r} + \delta U = 0$$

[The term 'virtual' is used because the displacements etc. are not necessarily physically realized (Langhaar 1962, 13).] Fraser reports that from **MMPVW**, Lagrange infers his equations of motion (Fraser 1985, 173 on which I lean). Lagrange called the **PVW**, the **principle of virtual velocities** in (Lagrange 1997). There he also privileges it by crowning it with the status of "a kind of axiom of mechanics" (Lagrange 1997, 26). Notice how central forces are in all of the above.

[61] (Hankins 1980, 184).

where $f(r_{ij})$ here represents a force law specifying details describing a repulsive or attractive force.[62] The potential function therefore represents operating forces in SYS because it represents work. That is why Hamilton calls it the "force-function". This is how forces enter Hamilton's approach to classical dynamics even after Hamilton introduces his auxiliary function $S$ around which (Hamilton, Second Essay 1835), is centered. For when Hamilton transforms the characteristic function $V$ into the auxiliary function $S$, he specifies it thus:

$$S = V - t\mathcal{H}$$

where $S$ is a function of $x, y, z, a, b, c$, and $t$, whose variation is:

**(Transformed LVA (T-LVA))**:

$$\delta S = -\mathcal{H}dt + \sum_{i=1}^{n} m_i \left( (\dot{x}_i \delta x_i - \dot{a}_i \delta a_i) + (\dot{y}_i \delta y_i - \dot{b}_i \delta b_i) + (\dot{z}_i \delta z_i - \dot{c}_i \delta c_i) \right)$$

Therefore, Hamilton transformed the **LVA** into the **(T-LVA)** using $S$ (see proposition W[7] in (Hamilton 1834, 307); (*qq.v.*, n. 63 and n. 64). This new auxiliary function $S$, which in (Hamilton, Second Essay 1835, 95) is called the "principal function", harbors within both $\mathcal{H}$ and $V$, and thereby includes the force-function $U$. It too must satisfy two fundamental PDEs that are really specifications of the force-function at the beginning (time $t_0$) and end of an evolution[63]:

**(New Constraining Fundamental PDEs (or the Hamilton-Jacobi Equation(s)))**:

$$\begin{cases} \dfrac{\partial S}{\delta t} + \sum\limits_{i=1}^{n} \dfrac{1}{2m_i} \left\{ \left(\dfrac{\partial S}{\partial a_i}\right)^2 + \left(\dfrac{\partial S}{\partial b_i}\right)^2 + \left(\dfrac{\partial S}{\partial c_i}\right)^2 \right\} = U_{t_0} \\ \dfrac{\partial S}{\partial t} + \sum\limits_{i=1}^{n} \dfrac{1}{2m_i} \left\{ \left(\dfrac{\partial S}{\partial x_i}\right)^2 + \left(\dfrac{\partial S}{\partial y_i}\right)^2 + \left(\dfrac{\partial S}{\partial z_i}\right)^2 \right\} = U \end{cases}$$

In (Hamilton, Second Essay 1835), Hamilton defines $S$ *via* an integral:

$$S = \int_0^t (T + U)\, dt = \int_0^t \mathcal{L}\, dt$$

where $\mathcal{L}$ is the Lagrangian within which the force-function (*inter alia*) resides (Hankins 1980, 192; Lützen 1995, 10). This is the **principle of least action**. Hamilton derives this principle from the **LVA** (Hankins 1980, 186). The **LVA** is therefore upstream from the principle of least action. Hamilton also uses the principal function $S$ to state what has become known as **Hamilton's principle**[64]:

---

(**Hamilton's Principle (HP)**):

$$\delta S = \delta \int_0^t \mathcal{L} \, dt = 0$$

The **principle of least action** and the **HP** are parts of both modern Hamiltonian and Lagrangian mechanics.[65]

In (Hamilton 1834), Hamilton used the force-function to present his equations of motion (the **DPVW** is an equation of motion). Hamilton's canonical equations of motion in (Hamilton, Second Essay 1835) likewise use the force-function.[66] Moreover, for systems like SYS, modern Hamiltonian mechanics specifies the Hamiltonian in terms of the force-function (see **sect. 3**), and for some systems unlike SYS, the Hamiltonian is ordinarily specified in part in terms of the Lagrangian which (again) harbors $U$.

### 4.2    From Potential Energy to Forces

I have shown that the force-function was *used* by Hamilton to present central formulas of his dynamics, to state the Hamiltonian $\mathcal{H}$, and to state the principal function that is $S$. I have demonstrated that both the force-function for the initial time of an evolution of a physical system, as well as the force-function for the end-time of that same evolution, are used to provide constraints on the principal function $S$.[67] I have described in what way the force-function is *used* to present modern Hamiltonian mechanics. Furthermore, I have argued that *Hamilton* interpreted the force-function in such a way that it represents work and so also forces. (That forces were for Hamilton, causes of motion, is a fact established in **sect. 2**) But why think that forces play any indispensable role in correctly interpreting potential energy in *modern* Hamiltonian dynamics?

In our modern context (as during Hamilton's time), energy is but a measure of a physical system's power or ability to bring about work.[68] Work, however, is force multiplied by displacement.[69] Thus, if this characterization is correct, forces are essential to the interpretation of energy in general.

The following expresses a law of classical mechanics, whether Lagrangian, Newtonian, or Hamiltonian:

(**Potential Energy Identity (PEI)**):

$$U(\mathbf{r}) \equiv - \int_{\mathbf{r}_0}^{\mathbf{r}} \mathbf{F}(\mathbf{r}') \cdot d\mathbf{r}'$$

when the **r**s are position vectors, and the triple bar expresses a mathematical identity (the flanking expressions are equal no matter the values we consistently attribute to the variables). **PEI** says that if our choice point mass travels from an initial position $\mathbf{r}_0$ to another position $\mathbf{r}$ (*e.g.*, along a curve)**,** $U(\mathbf{r})$ will be minus the work done by the force $\mathbf{F}$ during the evolution. **PEI** is more than a law of classical mechanics. It is the definition of potential energy for the relevant system-types.[70] The work involved is virtual if the referenced displacement due to force is nonactual or unrealized. This should not detract from the view I intend to promote, for even if the work is virtual, the definition that is **PEI** underwrites the position that the concept of potential energy has content *about* work and so about forces (assuming a mental representation theory of concepts). I call this understanding or definition of potential energy in classical mechanics **the orthodox interpretation of potential energy**.

---

[65] See (Feynman, Leighton and Sands 2010, 19-8); (Taylor 2005, 239).

[66] (Nakane and Fraser 2002, 163).

[67] See the **New Constraining Fundamental PDEs.**

[68] The *Oxford Dictionary of Physics* defines energy as "[a] measure of a system's ability to do work" (Rennie 2015, 180).

[69] When there are disagreeing directions between force and displacement, the relevant equation becomes: $W = \mathbf{F} \cdot s \, \cos\theta$ (where $s$ is displacement).

### 4.2.1 Forces Beyond the Newtonian Formulation

Have I reverted to the Newtonian formulation of classical mechanics by explicitly connecting the potential energy function to a force? Not at all. You will recall that Hamilton (1834) forged the same type of connection when he affirmed **FF**. In addition, I remind the reader that Hamilton called the potential energy function the force-function. Even today, we think of potential energy as the "energy of interaction".[71] It should be no surprise then that forces remain part of the ontology of modern Hamiltonian mechanics. There is nothing distinctively anti-Hamiltonian about them. When a point mass is under sway of a central force field, modern Hamiltonian mechanics specifies the involved radial force (in polar coordinates) as follows[72]:

---

[70] See, *e.g.*, (Taylor 2005, 111) who says that "[w]e define $U(\mathbf{r})$" in terms of, $-W(\mathbf{r}_0 \to \mathbf{r})$. But,

$$-W(\mathbf{r}_0 \to \mathbf{r}) \equiv -\int_{\mathbf{r}_0}^{\mathbf{r}} \mathbf{F}(\mathbf{r}') \cdot d\mathbf{r}'$$

where $W$ is the work function. That we can define $U(\mathbf{r})$ in terms of work is enough for my purposes because, again, work has to do with force. Hecht remarked,

"…only changes in $PE$ are defined, and these are defined as the work done on a system by conservative forces. Such work is measurable as it is being done. However, once work is done it no longer exists and is no longer measurable; if $\Delta PE$ is *only* defined by work done (e.g., $mgh$, or $1/2kx^2$, or $1/2CV^2$) it cannot be measured in stasis while it supposedly exists." (Hecht 2019a, 500 emphasis in the original)

Some maintain that potential energy and work are identical. Coopersmith stated, "[p]otential energy and work were eventually seen to be one and the same (an integration of force over distance)" (Coopersmith 2015, 115).

It is common to understand the work quantity in such a way that it is deemed more fundamental than potential energy. In his classic graduate level text on classical mechanics, Cornelius Lanczos (1893-1974) stated,

"[t]he really fundamental quantity of analytical mechanics is not the potential energy but the work function…In all cases where we mention the potential energy, it is tacitly assumed that the work function has the special form $W = W(q_1, q_2, ..., q_n)$, together with the connection $U = -W$" (Lanczos 1970, 34)

I have rephrased Lanczos's equations by using my own notation. The first equation in the quotation is inserted here in place of Lanczos's reference to equation (17.6). I argue that forces are more fundamental than potential energy in **sect. 4.3** below. Lanczos argues that work is more fundamental than force (Lanczos 1970, 27). I disagree. Work is analyzed in terms of force.

[71] (Coopersmith 2015, 337).

[72] (Taylor 2005, 531-532).

$$\mathbf{F}_{radial} = -\frac{dU}{dr}$$

What one should remember is that in many different contexts, Hamiltonian mechanics is correctly used to arrive at the same precise equations one can discover using force-laden Newtonian mechanics.[73] In other words, modern Hamiltonian mechanics does in fact provide important information about forces.

Some will accept that forces are central to modern Hamiltonian mechanics. At the same time, they will insist that for the general case of an $n$-particle system, Hamiltonian mathematical models use generalized forces and (quoting North) "[i]t isn't clear that they count as regular forces of the Newtonian kind".[74] A generalized (conservative) force with $n$-components is a higher dimensional vector that lives in a higher dimensional abstract mathematical space.[75] But in analytical mechanics of either the Lagrangian or Hamiltonian varieties, a generalized (conservative) force $Q_i$ depends upon and is asymmetrically determined by real-world (North's "Newtonian") forces whether due to either external fields or interactions between constituents of systems. Consider now my argument for this conclusion.

If there are real-world forces $\mathbf{F}_i$ acting between point-particles with masses given by $m_i$, we say that their total work is given by[76]:

(**Total Work or DPVW\***):

$$dW = \sum_{i=1}^{n}(X_i dx_i + Y_i dy_i + Z_i dz_i)$$

*assuming that* the particle constituents of the system have Cartesian coordinates that change by the arbitrary infinitesimal amounts represented by $dx_i, dy_i$, and $dz_i$, and that $X_i, Y_i$, and $Z_i$ give the components of the real-world forces $\mathbf{F}_i$ in appropriate orthogonal directions (*qq.v.*, n. 75, n. 76, and n. 78). A quick comparison reveals that this equation is equivalent to Hamilton's **DPVW** above. Modern analytical (*e.g.*, Hamiltonian) mechanics exploits the fact that the Cartesian coordinates are functions of the generalized coordinates $(q_1, …, q_n)$ such that the generalized coordinate variables have an invariant first-order differential form that is linear[77] and expressed by[78]:

$$dW = (F_1 dq_1 + F_2 dq_2 + F_3 dq_3 … etc … F_n dx_n)$$

---

[73] (ibid., 532).

[74] (North forthcoming, 14).

[75] I will assume that generalized force $Q_i$ has the dimension of a force and not that of the moment of a force. On this distinction, see (Langhaar 1962, 17). For more on generalized force in general, see (Fitzpatrick 2011); (Lanczos 1970, 27-31); (Langhaar 1962, 14-23); (Peacock and Hadjiconstantinou 2007); (Sommerfeld vol. 1 1964, 187-189); (Stewart 2016, 16-21). In places, my discussion follows these sources.

[76] Following (Lanczos 1970, 28).

[77] (Lanczos 1970, 28).

[78] We could use $\delta_n$ instead of $d_n$ to highlight the fact that the involved force impression can be virtual (as in Langhaar 1962, 16).

The coefficients $F_1, F_2, F_3$, etc. are not the components of the real-world force $\mathbf{F}_i$, but are instead the components of generalized force $Q_i$, that higher dimensional abstract vector previously mentioned (more on these components soon). I will now use the symbol: $\mathcal{F}_i$ to represent the generalized force components.

$Q_i$ can be thought of as an abstract actor in the $n$-dimensional configuration space used to model the system. Points no longer just travel or orbit without reason, they orbit *because* the generalized force with its $n$-components "acts", "producing" the orbit (Lanczos 1970, 28-29). Of course, this is not an action indicative of causation. The 'because' involved here is one "without cause". The explanation provided is an explanation by constraint as in (Lange 2017). However, we have reason to believe that these explanations by constraint in the abstract configuration or phase spaces are determined by causal explanations in the real world.

Consider that the components of $Q_i$ can be related to potential energy and generalized coordinates as follows:

$$\mathcal{F}_i = -\frac{\partial U}{\partial q_i}$$

But by **PEI**, we know that one can also state the generalized force components in terms of work and generalized coordinates:

$$\mathcal{F}_i = \frac{\partial W}{\partial q_i}$$

What these equations reveal is that one can calculate generalized (conservative) forces from the potential energy or work functions. This is because the real-world forces lying beneath work (work is force times displacement) and potential energy (identical to minus work) determine generalized force (*q.v.*, **GF** below). But the minus partial derivatives of potential energy give the components of real-world forces (*q.v.*, **sect. 4.2.2** below). Of course, matters are not resolved. When the work function is time-dependent, there is no potential energy to look to for the purposes of calculating generalized force. Of course, this bothers my project none because work remains, and residing in work is real-world force. Furthermore:

**(Generalized Force (GF))**:

$$Q_j \equiv \sum_{i=1}^{n} \mathbf{F}_i \cdot \frac{\partial \mathbf{r}_i}{\partial q_j} \ (j = 1, \dots, m)$$

where here, $\mathbf{r}_i$ gives the position vector for the $i^{\text{th}}$ point where a *real-world force* is applied (assuming that $\mathbf{r}_i$ is a function of the generalized coordinates) such that its variation is given by[79]:

$$\delta \mathbf{r}_i = \sum_{j=1}^{m} \frac{\partial \mathbf{r}_i}{\partial q_j} \delta q_j$$

---

[79] See (Fitzpatrick 2011); (Peacock and Hadjiconstantinou 2007, 6); (Stewart 2016, 19).

Most importantly, $\mathbf{F}_i$ represents real-world forces such as gravitation. A plain implication of all of this (especially **GF**) is that generalized force depends upon real-world applied forces. Indeed, the components of $Q_j$ represent the numerous forces standing behind $\mathbf{F}_i$, *i.e.*, the real-world forces at work in the system. Thus, generalized force $Q_j$ is a more abstract representation of force impressions and actions in the world. That is why Lanczos (1970, 29) says, "the dynamical action of all the forces can…be represented by a single vector [*i.e.*, the generalized force vector] acting on" a point in a higher-dimensional mathematical space.

### 4.2.2 Back to Potential Energy

Some reject the orthodox interpretation of potential energy, choosing instead to identify potential energy with a "capacity to do work".[80] Call this outlook **the capacity interpretation of potential energy**. On this view, potential energy just is an *objective* dispositional property of physical systems.

While the capacity interpretation is not identical to the orthodox approach, it is nonetheless friendly to my central thesis because it causally interprets the potential energy function by ascribing to choice systems with potential energies causal powers, *viz.*, capacities or dispositions to act by force impression over a distance thereby performing work. Ignoring for now the benefits this account would provide for my cause, I should detail how its fatal flaw follows from a point made by some of the very promulgaters of the capacity interpretation (*i.e.*, Marion and Thornton 1988, 73). If the capacity interpretation were correct, whether a system has an *objective* potential energy or not would be a fact insensitive to conventions and arbitrariness introduced by physicists. Howbeit, there is no means whereby one can measure such an objective potential energy that is an objective capacity.[81] The quantity physicists call "potential energy" in the three main formulations of classical mechanics is a quantity whose differences alone ultimately matter because the differences are what can be measured. Of course, the fact that a system's evolution over time can be described by appeal to differences in potential energy presupposes that it at least makes sense to ascribe to said system potential energy full stop (no differences). However, the problem is that those full stop attributions are arbitrary. In order to get the differences physics needs, physicists predicate potential energies to systems at locations in an arbitrary manner. That is why the differences are emphasized.

Consider the following well-known and often repeated example used to help teach students about potential energy. Let spatial point $x$ at height $h(x)$ be the summit of Mauna Kea (the tallest

---

[80] (Marion and Thornton 1988, 72).

[81] What is said in the main text holds true for both internal and external potential energy. None "of these potential energies is independently measurable" (Hecht 2016, 10). What I say here also holds true for static electric field energy (Hecht 2019b, 3) (*q.v.*, **sect. 5**).

Albert Einstein (1879-1955) said,

> "[b]ut if every gram of material contains this tremendous energy, why did it go so long unnoticed? The answer is simple enough: so long as none of the energy is given off externally, it cannot be observed. It is as though a man who is fabulously rich should never spend or give away a cent; no one could tell how rich he was." (Einstein 1954, 340)

At this point, I should acknowledge my indebtedness to the work of Eugene Hecht (2016, 2019a, 2019b and *q.v.*, n. 96). His papers cite several of the sources I cite in this work.

mountain on earth measured from base to summit). Put a simple particle system or body $b$ there. One now states the (gravitational) potential energy of the system as follows:

**(Gravitational Potential Energy (Approx.) (GPEA))**:
$$U_g = mgh(x)$$

But here $g$ is gravitational acceleration due to an acting gravitational force[82], $m$ is the gravitational mass (equivalent to the inertial mass) of a body $b$ at that height, and $mgh(x)$ is interpreted as the *work* required to get $b$ to $h(x)$ given that $b$ started somewhere else, *e.g.*, sea level (you will notice that work is now being set equal to the gravitational potential energy) without imparting kinetic energy to $b$ while laboring against gravity. But sea level here was chosen arbitrarily to facilitate acquisition of the differences one needs to make sense of $U_g$.[83] Sea level is arbitrarily determined to be the place where $U_g$ vanishes (French 1971, 378-379). How then can the potential energy referenced here be an objective property of the system when its quality and quantity are determined by the arbitrary decisions of physicists? Of course, this arbitrariness will infect the quantity that is (virtual) work too, but the infection is not overly contagious. The $mg$ portion (weight) of **GPEA** expresses the force pulling the object toward the center of the earth.[84] The only way to get $b$ to $h(x)$ without giving it kinetic energy is to apply an external force to $b$ (French 1971, 378). When any conservative force acts thereby producing an acceleration, what transpires is a purely objective affair. More reasons will be given for this judgment soon (*q.v.*, **sect. 4.3**).

That the most important points in the preceding discussion of this subsection (*viz.*, those about (a) arbitrariness and potential energy in general, (b) arbitrariness and gravitational potential energy in particular, and (c) the **GPEA**) hold true in both Lagrangian and Hamiltonian formulations of classical mechanics is well-known.[85] There is therefore no escape from my reasoning by appeal to a shift in formalism. But we can say something more general. In almost every context, the specification of the potential energy function does not change across the three major formulations of classical mechanics.

Let us now entertain an objection to my argument from arbitrariness. The **GPEA** is but an approximation even in classical mechanics. It holds true for bodies near to the earth, but as one moves away from the earth, one will need to use a different expression for gravitational potential energy, one that follows from Newton's famous universal law of gravitation and the *definition* that is **PEI**:

**(Gravitational Potential Energy (GPE))**:

$$U_g = -\frac{GMm}{r}$$

---

[82] Sometimes it is interpreted as the gravitational force field.

[83] "Note that we cannot measure the absolute potential energy, but only differences." (Ryder 2007, 58). *Absolute potential energy does not fail to make sense because it is not measurable. There is no verificationism afoot here.*

[84] It is therefore no surprise that the SI unit of measurement for both weight and force is one and the same, *viz.*, the newton.

[85] (Fecko 2006, 517-518); (Penrose 2007, 475); (Pletser 2018, 55); *cf.* (Taylor 2005, 542 eq. 13.51). The Hamiltonian would be similarly specified in the Hamilton-Jacobi formulation as well. See (Pletser 2018, 55); (Torres del Castillo 2018, 256, 266-267).

where $M$ is the gravitational mass of the earth (the body exerting the gravitational force), $m$ is the gravitational mass of body $b$, $r$ is the distance between $b$ and the earth, and big $G$ is the gravitational constant.

Switching to **GPE** does not help matters. That law is usually posited given yet another arbitrary choice. This time it is to set the gravitational potential energy of $b$ that is $U_g$ equal to zero when $b$ is situated at a position infinitely far removed from the earth. Potential energy decreases as $b$ moves closer to the earth and so is negative.[86] Physicists will usually add that this arbitrary decision is reasonable, natural, or intuitive, but I have been unable to find an argument for such a choice that would render that choice principled or well-justified.[87] I do not question that it is convenient for calculations, but that does not mean it informs us about what the world is like even given a robust scientific realism. There are fictional devices aplenty in physics that help us with our calculations, but no one would commit to the existence of these devices. Who likes Gaussian surfaces?[88] Potential energy, yes, even gravitational potential energy, can only be specified up to an *arbitrary* additive constant. The arbitrariness is unavoidable.

Let us now consider a second reason for associating forces with potential energy. The orthodox interpretation of potential energy is provided by **PEI**, though it may need some tinkering in order to handle various other system-types. I maintain that because orthodoxy defines potential energy in terms of work, and work in terms of forces, causation enters mechanics through the potential energy function. Interestingly, my analysis of matters can be found in some of the very earliest work in which potential energy was first used in physics. For example, William Rankine (1820-1872; he coined the term 'potential energy' in 1853) remarked,

> In this investigation the term *energy* is used to comprehend every affection of substances which constitutes or is commensurable with a power of producing change in opposition to resistance, and includes ordinary motion and mechanical power, chemical action, heat, light, electricity, magnetism, and all other powers, known or unknown, which are convertible or commensurable with these. All conceivable forms of energy may be distinguished into two kinds; actual or sensible and potential or latent…*Potential energy*…is measured by the amount of a change in the condition of a substance, and that of the tendency or force whereby that change is produced (or, what is the same thing, of the resistance overcome in producing it), taken jointly.[89]

Rankine is here associating energy with productive power (causation), but he is also associating potential energy with work and therefore also forces. Eight years later, William Thomson's (or Lord Kelvin's) discussion of the electric potential function (which is used in classical electrostatics to state the potential energy function (*q.v.*, **sect. 5** below)) strongly associated that potential with work. He said,

---

[86] All of this remains as I am presenting it in both Newtonian and Hamiltonian mechanics. See (Meyer and Offin 2017, 61-62) for a discussion of gravitation and classical non-relativistic Hamiltonian mechanics.

[87] I should add that I believe that the arbitrariness is to blame for the negativity. The arbitrariness is also the reason why negative gravitational potential energies should not confound the metaphysician of physics (Mann 2018, 17). If energies are but measures or convenient ways of representing what's really happening with forces, then the signs of the convenient devices need not bother one. It's what's fundamental that matters.

[88] The reference to Gaussian surfaces wasn't just for the purposes of being humorous. Gaussian surfaces are *arbitrarily* specified closed surfaces introduced by the physicist to help with calculations in electrodynamics.

[89] (Rankine 1853, 106) emphasis in the original.

*Electric potential*. —The amount of work required to move a unit of electricity from any one position to any other position, is equal to the excess of the electric potential of the second position above the electric potential of the first position.[90]

Thomson here says that differences are what matter, and that the electrostatic potential (and so the electrostatic potential energy) just is work (whether virtual or not) required to complete a task.

Both Rankine and Thomson's views of energy are important because they influenced the work of Rudolf Clausius (1822-1888), James Clerk Maxwell, and Ludwig Boltzmann (1844-1906). These mechanicians thought of entropy as a quantity that tracks how the energy (as understood by Rankine and Thomson) transforms over time.[91] Modern thermodynamics and statistical mechanics has thereby inherited the work-laden and so also force-laden notion of energy. It is therefore unsurprising to see in the work of modern thermodynamicists, such as Klein and Nellis, the following: "…the property entropy is introduced in order to quantify the quality of energy"[92] and "[t]he Second Law states that the quality of energy, i.e., the capability to do work, is reduced in all real processes."[93] All of this is unsurprising on the truth of the orthodox interpretation of potential energy.

That energy and work should be strongly associated is evidenced by the further fact that the SI derived unit of *both* energy *and* work is the joule **J**. One joule just is a unit of measurement about a force, more specifically, the work performed or executed by one newton (**N**) of force over one meter (**m**) in a single direction, *viz.*, the direction of the force impressed. In the CGS (centimeter-gram-second or the Gaussian) unit system, the unit of *both* energy *and* work is the *erg*. The abbreviation *erg* derives its meaning from the Greek ἔργον, which means work. Therefore, it is unsurprising that the *erg*, like the joule, is a unit of measurement about a force, more specifically, the work performed or executed by one dyne (**dyn**; the unit of force in the CGS system) of force over one centimeter (**cm**; the unit of length in the CGS system) in a single direction, *viz.*, the direction of the force impressed. These facts are not at all surprising, given the truth of the orthodox interpretation of potential energy. It is what you'd expect.

My final line of evidence has (I believe) some serious bite. To feel its teeth, simplify matters some by giving attention to a target system of but one point mass that travels along a curve in three dimensions of space due to a conservative force. Let us call this new target system SYS-1. We will assume that SYS-1 is conservative and holonomic. In modern Hamiltonian mechanics, the force in SYS-1 will be, what is sometimes called the **corresponding force of the potential energy** function used to represent the evolution of SYS-1. This corresponding force in SYS-1, is the referent of **F** in the integral statement of the potential energy function that is **PEI** stated above. The relation of correspondence with respect to SYS-1 can be stated quantitatively as:

$$\mathbf{F} = -\nabla U$$

---

This equation says that the minus partial derivatives of the potential energy function express the components of the corresponding force in particular directions.[94] It should now be clear precisely how potential energy is intimately related to corresponding forces.

### 4.3    Forces are More Fundamental than Potential Energy

There are those who wonder whether there is any reality to potential energy at all.[95] I will not go so far as to challenge the existence of potential energy. I will instead argue that potential energy is less fundamental than force.[96] This conclusion seems to be in line with a well-represented position among physicists.[97]

My first argument for the view that potential energy is less fundamental than force is as follows. With respect to conservative physical systems, facts about potential energies are derivable from facts about conservative forces. Likewise, facts about conservative forces are derivable from facts about potential energies. However, nonconservative force-facts (about nonconservative systems) are not derivable from facts about potential energies. Nonconservative forces can act when no potential energies are specifiable (indeed when no potential energy exists). This constitutes evidence for the claim that potential energy cannot exist in the absence of conservative forces, and that forces can exist in the absence of potential energy. Thus, if one depends upon the other, forces serves as the dependency base, not potential energy.

Second, we have, in essence, already provided a reason for regarding forces as more fundamental than potential energy. The latter quantity has no absolute or objective status. One can only specify it up to an *arbitrary* additive constant. In order to evaluate facts about potential energy, one must admit arbitrariness, and even after the necessary arbitrariness is in play, differences in potential energies are all that one can have access to. By contrast, when a force is impressed thereby producing an acceleration, that causal fact will hold relative to every inertial frame of reference in Hamiltonian mechanics. Conservative forces are Galilean invariant quantities, and the equations of motion which relate those forces to resulting motions are Galilean covariant.[98] However, potential energy is not Galilean invariant.[99] There is no guarantee that

---

[94] *E.g.*, $-\frac{\partial U}{\partial z}$ expresses the component of force along the direction of the $z$-axis in the rectangular coordinate system (Goldstine 1980, 176); (Taylor 2005, 111, 116-117).

[95] (Sullivan 1934, 247-248).

[96] I believe this is the best philosophically sophisticated characterization of the position defended in (Hecht 2003, 2007, 2016, 2019a, 2019b).

[97] As Robert Mills (of Yang-*Mills* theory) said, "the idea of potential energy is not truly fundamental and that it breaks down in the relativistic world…" (Mills 1994, 152) *q.v.*, n. 81. *Cf.* (Lanczos 1970, 34) already quoted; and (Thomson 1888, 15). Coopersmith (2015, 339) argues that because kinetic energy has the same form for every system and potential energy does not, the former is more fundamental than the latter (citing (Maxwell 1871, 282) in support).

[98] (Raju 1994, 64). See also (Maudlin 2011, 174).

[99] As Coopersmith remarked in correspondence:

> "Even outside of Einstein's Relativity theories, it is not always true that the potential energy is invariant. A requirement for Galilean invariance is that the potential energy depends only on 'relative' coordinates (e.g. the difference between two positions) and not on 'absolute' coordinates (the position of the old oak tree at the corner of the street)." (11/08/2020)

See (Diaz et. al. 2009, 271-272) who remarked,

performing a Galilean transformation on the mathematical statement of potential energy for a system in an inertial frame will yield an equation of the same form preserving the same value for the potential energy of that system. We should deem invariant quantities more fundamental than non-invariant ones. At a non-relativistic classical world, we should deem Galilean covariant laws more fundamental than non-covariant laws. Is this not a lesson taught by the successes of our best physical theories? The laws of classical mechanics that have passed empirical muster are precisely those laws that are appropriately constrained by meta-laws that are symmetry or invariance principles.[100]

Robert Nozick (1938-2002) once said,

> Amalie Emmy Noether showed that for each symmetry/invariance that satisfies a Lie group, there is some quantity that is conserved. Corresponding to invariance under translation in space, momentum is conserved…and to invariance of the law under the addition of an *arbitrary constant* to the phase of the wave function, apparently electrical charge is conserved. So it is not surprising that laws that are invariant under various transformations are held to be more objective. Such laws correspond to a quantity that is conserved, and something whose amount in this universe cannot be altered, diminished, or augmented should count as (at least tied for being) the most objective thing there actually is.[101]

My scholarship is shaky. I have cut out that portion of the quotation of Nozick which reads: "Corresponding…to invariance under translation in time, [mechanical] energy is conserved" (Nozick 2001, 81). I have done this to emphasize that Nozick's invitation to his reader is for that reader to conclude that mechanical *energy* is one of "the most objective" things "there actually is."

---

> "…we illustrated that after a change of reference frame, the work done by each force also changes (even if the transformation is Galilean). Consequently, the corresponding potential energies change when they exist." (ibid., 272)

These authors give an argument for their conclusions at (ibid., 271-272).

Horzela et. al. (1991) stated that,

> "The explicit expressions of the potential energy as functions of the position $\vec{x}(t)$ all have noncovariant meaning and therefore may be valid only in one inertial reference frame." (ibid., 12, their argument for this starts on page 11)

With respect to force and acceleration, Tefft and Tefft (2007) state,

> "…quantities, such as acceleration and force, are invariant or, to use Newton's term, 'absolute' between inertial reference frames. Such quantities have the same values in any inertial frame." (ibid., 220)

[100] *Cf.* the discussion in (Earman 2004, 1230).

[101] (Nozick 2001, 81). John Earman (2004) likes the type of inference invited by Nozick in the context of classical mechanics, but believes it suffers important setbacks when dealing with general relativity.

> "The implementation of part of Nozick's formula *objectivity = invariance* by means of the constrained Hamiltonian formalism goes swimmingly: in case after case it yields intuitively satisfying results. But the application to Einstein's GTR yields some surprising and seemingly unpalatable consequences." (ibid., 1234)

On the contrary, while it is true that for isolated systems featuring only conservative forces total mechanical energy is conserved, that fact does not entail the Galilean invariance of total mechanical energy, potential energy, or kinetic energy. None of these quantities are Galilean invariant. That is why neither the Hamiltonian nor the Lagrangian are either.[102] However, for the types of systems with which I have been concerned, the Hamiltonian becomes total mechanical energy which, again, *is conserved*. Thus, not all conserved quantities are Galilean invariant quantities. And so, as Schroeren put matters, "Noether's theorem concerns the way particles behave *under temporal evolution*, i.e., whether certain physical quantities are conserved"; it does not concern whether those quantities are Galilean invariant.[103] Yet, Nozick was supposed to help us commit to Paul Dirac's (1902-1984) dictum that "[t]he important things in the world appear as the invariants…"[104] *These* important things are the objective things. At a non-relativistic classical world, it is Galilean *invariance* (using Nozick's wording) that "is [or should be]…connected to something's being an objective fact".[105] But while Hamilton's laws or equations of motion are covariant under Galilean transformations, as is the work-energy theorem, both those laws and that theorem fail to "correspond to a quantity that is conserved" and that is one of "the most objective thing[s] there actually is", *if* the quantity in question is taken to be the Lagrangian, the Hamiltonian,

---

[102] As Coopersmith stated, "energy is not an invariant quantity" (2015, 342). The Hamiltonian is not invariant under boost operations (Butterfield 2007, 6). Butterfield said, "the Hamiltonian of a free particle is just its kinetic energy, which can be made zero by transforming to the particle's rest frame; *i.e.* it is not invariant under boosts." (ibid.) This is true even in quantum mechanics (Lombardi et. al. 2010, 99). In classical mechanics, the Lagrangian is invariant under rotations and translations, but not under boosts (Finkelstein 1973, 106-107). Be careful. Landau and Lifshitz's famous text on mechanics says that "the Lagrangian is [Galilean] invariant", but their argument only demonstrates that Lagrange's equations of motion are invariant under Galilean transformations (Landau and Lifshitz 1976, 7).

Interestingly, Coopersmith (2015, 239) says that the "potential energy function" just "is the Hamiltonian…and it is the function that determines the entire dynamics of the system." If Coopersmith is right, it would not be surprising that potential energy fails to be a Galilean invariant quantity (*cf.*, ibid., 313).

[103] (Schroeren 2020, 52) emphasis in the original. Schroeren goes on to correctly note how "every property linked to a symmetry in the relevant sense is invariant under that symmetry" (ibid.). So, some type of thesis regarding the invariance of energy may be saved. However, that type of invariance cannot be indicative of that which is one of "the most objective thing[s] there actually is" for, again, total mechanical energy is conserved but not objective at least because of the arbitrariness that sneaks into potential energy (*q.v.*, **sect. 4.2**). I do not know if Schroeren would agree with my conclusions.

[104] (Dirac 1995, 456). Nozick himself uses this quotation at (Nozick 2001, 76).

[105] (Nozick 2001, 76).

Someone may ask: But what about the action that is minimized in Lagrangian and Hamiltonian mechanics? That quantity *is* invariant under Galilean transformations, and it is typically understood in terms of energy multiplied by time. I reply that it is typically the Lagrangian multiplied by a small change in time (usually flanked by the integration symbol (the "action integral")). It is difficult to discern the metaphysical nature of that quantity, but it is far from potential energy alone. My view of the relationship between force and action belonged to both Euler and Lagrange. The action and action integral track the evolution of the system by indirectly representing its dynamical force interactions. That is why when you shift from one system with an operating force **F_n** to a system with different operating forces, the form of the action changes. Euler recognized that if the system involves accelerations, the action is a minimum, ***given*** that the system being modeled is acted upon by forces (Euler 1744, 311-312). The correct act of integration yields the system's trajectory only under the assumption that a force has acted and that the form of the integral is appropriately specified in light of that force-action. This same point was made by Lagrange. See (Lagrange 1867, 365-468) and the translated quotation at (Boissonnade and Vagliente 1997, xxxiii). So, the explanation of the invariance of the action integral arises from the invariance of the acting forces, those same forces that the form of the action integral is sensitive to.

potential energy, kinetic energy, or even work.[106] Nozick's inference runs from invariance or symmetry principles (perhaps together with or closely followed by empirical data), to invariant (or covariant) laws, to conserved quantities, and then to objective quantities. But that inference is a bad one. Nozick's inference should have had the following structure for the non-relativistic classical case:

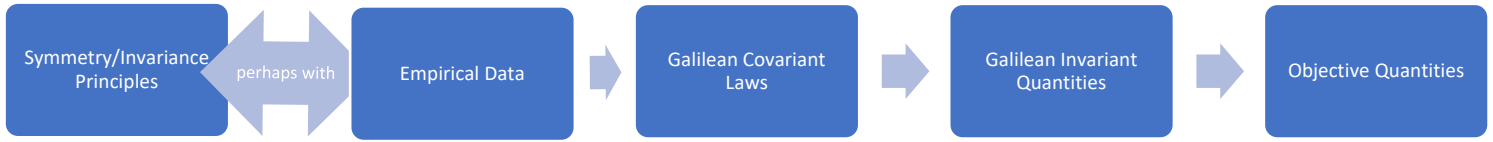

What explains the Galilean covariance of the laws of classical mechanics (including Hamilton's equations of motion) is the nature of operating forces. That is why Tim Maudlin was right to stress that "…the necessary and sufficient condition for any Newtonian [classical but non-relativistic] theory to be Galilean invariant is that the force **F** be the same in all inertial frames."[107] Forces *are* invariant and objective. So too are their resulting accelerations. Potential energy is neither invariant nor objective. The objective things are more fundamental than the non-objective things. Thus, force is therefore more fundamental than potential energy.

## 5    Potential Energy and Classical Electrostatics

There exists a weighty analogical consideration in favor of the claim that potential energy in classical mechanics is grounded in forces at work in the world. The argument starts out by referencing the known marked similarities between, on the one side, potential energy $U_E$ in electrostatics and the electrostatic potential $\Phi$ (in that same science), and on the other side, the potential energy function $U$ in classical mechanics.

Consider a conservative isolated system of classical electrostatics (SYS$_E$). Here the electric potential $\Phi$ "is the potential energy of a unit charge"[108] $q$, such that:

$$U_E = q\Phi$$

I have already shown how potential energy is related to work in mechanics. To show off that same relationship-type in electrostatics, add to SYS$_E$ an electric field **E**.[109] In electrostatics, this field will stand in the same relation to $\Phi$ that **F** stands in relation to $U$ in classical mechanics.

$$\mathbf{E} = -\nabla\Phi$$

Electric force becomes,

---

[106] The incorporation of work into the list should not be surprising. As I've already argued, energy should be analyzed in terms of work. Energy is not Galilean invariant, and so neither is work.

[107] (Maudlin 2011, 174) emphasis in the original.

[108] (Shankar 2016, 97).

[109] I assume that SYS$_E$ is one for which it is true that the curl of the electric field equals zero.

(**Simple Electric Force (SEF)**):

$$\mathbf{F}_E = \mathbf{E}q = -\nabla(q\Phi) = -\nabla U_E$$

If $\epsilon_0$ is the electric or vacuum permittivity constant, and there are two point charges $q_1$ and $q_2$, then the force between them (which will yield an equal and opposite force) will be given by Coulomb's law:

(**Coulomb's Law (CL)**)[110]:

$$\mathbf{F}_E = \frac{1}{4\pi\epsilon_0} q_1 q_2 \frac{\mathbf{r}_1 - \mathbf{r}_2}{|\mathbf{r}_1 - \mathbf{r}_2|^3}$$

**CL** says that the electric force is determined by the locations and properties of point charges. In electrostatics, the electric field is likewise determined by point charges. That is why one can motivate the following expression for the electric field by appeal to **CL** (*q.v.*, also n. 110):

(**Simple Electric Field (SEFi)**):

$$\mathbf{E}(\mathbf{r}) = \frac{1}{4\pi\epsilon_0} q_1 \frac{\mathbf{r} - \mathbf{r}_1}{|\mathbf{r} - \mathbf{r}_1|^3}$$

which specifies the electric field $\mathbf{E}$ at $\mathbf{r}$. This equation says that $\mathbf{E}$ is *generated* by point charge $q_1$.[111]

When a field $\mathbf{E}$ is in play, the electrostatic potential is reified, and one introduces to $\text{SYS}_E$ an array of $(n-1)$ charges $q_j$ situated in spatial region $\mathbf{r}_j$, that array will *generate* a potential $\Phi(\mathbf{r}_i)$, such that[112]:

(**Potential from Array (PFA)**):

$$\Phi(\mathbf{r}_i) = \frac{1}{4\pi\epsilon_0} \sum_{j=1}^{n-1} \frac{q_j}{|\mathbf{r}_i - \mathbf{r}_j|}$$

And this helps us support the judgment that the total potential energy of the system featuring the point charge array, electric field, and potential is:

(**Potential Energy and Work (P&W)**[113]):

$$W = \frac{1}{4\pi\epsilon_0} \sum_{i=1}^{n} \sum_{j<i} \frac{q_i q_j}{|\mathbf{r}_i - \mathbf{r}_j|}$$

---

[110] We can, of course, simplify many of these expressions (as in Shankar 2016, 19-41), but I like to explicitly state the relations as they appear in the main text thereby following the well-regarded exposition of electrostatics in (Jackson 1999, 24-56).

[111] (Jackson 1999, 24-25, 29).

[112] (Jackson 1999, 40).

[113] (Jackson 1999, 41); (Greiner 1998, 28).

Despite the fact that I've here specified the potential energy of the system (*i.e.*, the potential energy of every charge under the influence of every force in the system), my inclusion of the work variable $W$ on the left is purposeful. In this context when you give the work in electrostatics you give the potential energy. Once again, this is precisely what you'd expect if potential energy were correctly analyzed in terms of work. But why should one believe that forces wrought by the point charges are more fundamental than potential energy? Why should one believe that forces ground potential energy?

As with potential energy in classical mechanics, the potential energy of systems like SYS$_E$ can only be specified up to an arbitrary additive constant.[114] The same is true of the electrostatic potential. Thus, any specification of these quantities will involve some arbitrariness and it is because of this arbitrariness that orthodoxy in electrostatics typically treats the potential as a conventional and derivative device that helps with calculations. In David Griffiths' widely used textbook, we have: "[p]otential as such carries no real physical significance, for at any given point we can adjust its value at will by a suitable relocation of…[the reference point]".[115] Olness and Scalise remarked, "…the potential itself is not a physical quantity. In particular, we can shift the potential by a constant…and the physical quantities will be unchanged."[116] And while any good theory of electrostatics must be relativistic, both $U_E$ and $\Phi$ fail the test of Lorentz invariance (Steane 2012).

Orthodoxy in classical electrostatics has it that $U_E$ is closely connected to work and force. Equation **P&W** and my subsequent commentary makes that clear. And because $U_E = q\Phi$, the electrostatic potential is also strongly associated with both work and force. As Feynman said,

(a) the "electric potential…is related to the work done in carrying a charge from one point to another."[117]

But Feynman likewise expounded orthodoxy when he added that:

(b) "The existence of a potential, and the fact that the curl of **E** is zero, comes really only from the symmetry and direction of the electrostatic forces."[118]

---

[114] See the proof in (Langhaar 1961, 19)

[115] (Griffiths 2017, 81).

[116] (Olness and Scalise 2011, 309). In Maxwell's *An Elementary Treatise on Electricity*, he said that "[t]he electric potential…which is the analogue of temperature is a mere scientific concept. We have no reason to regard it as denoting a physical state" (Maxwell 1888, 53).

[117] (Feynman, Leighton and Sands 2010, 4-4).

[118] (Feynman, Leighton and Sands 2010, 4-7). The fact that the curl of the electric field equals zero, or:

$$\nabla \times \mathbf{E} = 0$$

which can be derived from the generalized Coulomb's law:

$$\mathbf{E}(\mathbf{r}) = \frac{1}{4\pi\epsilon_0} \int \rho(\mathbf{r}') \frac{\mathbf{r} - \mathbf{r}'}{|\mathbf{r} - \mathbf{r}'|^3} d^3 r'$$

where $d^3 r'$ gives the 3D volume element at $\mathbf{r}'$, and where $\rho(\mathbf{r}')$ gives the volume charge density at $\mathbf{r}'$ (a well-known fact mentioned in numerous places, but see ibid, 4-3, 4-7 and *q.v.*, n. 111). This derivation-fact supports Feynman's claim.

Orthodox tenets (a) and (b) are both plainly supported by the picture I have painted with the help of the preceding equations. $U_E$ and $\Phi$ are strongly related such that: $U_E = q\Phi$ (and **SEF**). The electric field and the electrostatic potential are strongly associated with one another, hence: $\mathbf{E} = -\nabla\Phi$ and **SEF**. But both $\mathbf{E}$ and $\Phi$ are determined by forces between point charges, hence the use of **CL** to motivate **SEFi** (*q.v.*, n. 118), and in addition (considered separately) the fact that the **PFA** holds.[119] **P&W** tells us that the work of a system depends on the positions and natures of point charges. The positions and natures of point charges determine the electric forces at work. This is not surprising because, again, physics teaches that work should be characterized in terms of forces. According to the **P&W**, that work just is the system's potential energy and so, orthodoxy in electrostatics says that "the potential energy…depends on the [acting] forces..."[120] And because there are so many known similarities between potential energy in electrostatics $U_E$ and potential energy $U$ in classical mechanics (some of which I have tried to highlight here), we can infer that it is likely that potential energy in classical mechanics is likewise downstream from and determined by forces.

## 6    Conclusion

I have shown how Hamilton's philosophical commitments led him to a causal interpretation of classical mechanics. I argued that Hamilton's metaphysics of causation was injected into his dynamics by way of a causal interpretation of the force quantity. I then detailed how forces remain indispensable to both Hamilton's formulation of classical mechanics and what we now call Hamiltonian mechanics (*i.e.*, the modern formulation). On this point, my efforts primarily consisted of showing that the orthodox interpretation of potential energy is none other than that interpretation found in Hamilton's work. Hamilton called the potential energy function the force-function because he believed that it represented forces at work in the world. Multifarious non-historical arguments for the orthodox interpretation of potential energy were provided, and matters were concluded by showing that in classical Hamiltonian mechanics, facts about the potential energies of systems are grounded in facts about forces. If one can tolerate the view that forces are causes of motions, then Hamilton provided the road map for transporting causation into one of the most mathematically sophisticated formulations of classical mechanics, *viz.*, modern Hamiltonian mechanics.

---

[119] Charges near conductors likewise create electrostatic potentials (Appel 2007, 165).
[120] (Shankar 2016, 82).